\newcommand*{\ARXIV}{}%
\ifdefined\ARXIV
\else
\fi

\ifdefined\ARXIV
\documentclass{article}
\usepackage{arxiv}

\else
\documentclass[sigconf, screen, nonacm]{acmart}
\fi

\AtBeginDocument{%
  \providecommand\BibTeX{{%
    \normalfont B\kern-0.5em{\scshape i\kern-0.25em b}\kern-0.8em\TeX}}}

\ifdefined\ARXIV
\else
\setcopyright{acmcopyright}
\copyrightyear{2021}
\acmYear{2021}
\acmDOI{10.1145/1122445.1122456}

\acmConference[ACM'21]{ACM '21: ACM Symposium on ACM Symposiums}{June 03--05, 2021}{Online, Earth}
\acmBooktitle{ACM '21: ACM Symposium on ACM Symposiums, June 03--05, 2021, Online, Earth}
\acmPrice{15.00}
\acmISBN{978-1-4503-XXXX-X/18/06}
\fi

\title{Effectiveness of Social Virtual Reality}

\ifdefined\ARXIV

\author{Lisa Izzouzi, Anthony Steed\\
Department of Computer Science\\University College London\\
\texttt{l.izzouzi@ucl.ac.uk, A.Steed@ucl.ac.uk}
}
\else
\author{Lisa Izzouzi, Anthony Steed}
\email{l.izzouzi@ucl.ac.uk, A.Steed@ucl.ac.uk}
\orcid{0000-0001-9034-3020}
\affiliation{%
  \institution{Department of Computer Science, University College London}
  \streetaddress{66-72 Gower St}
  \city{London}
  \country{UK}
  \postcode{WC1E 6EA}
}

\fi

\ifdefined\ARXIV
\else

\ccsdesc[500]{Computing methodologies~Virtual reality}
\ccsdesc[500]{Human- centered computing~Virtual reality}
\fi

\begin{document}

\ifdefined\ARXIV
\maketitle
\else

\fi

\begin{abstract}
A lot of work in social virtual reality, including our own group's, has focused on effectiveness of specific social behaviours such as eye-gaze, turn taking, gestures and other verbal and non-verbal cues. We have built upon these to look at emergent phenomena such as co-presence, leadership and trust. These give us good information about the usability issues of specific social VR systems, but they don't give us much information about the requirements for such systems going forward. In this short paper we discuss how we are broadening the scope of our work on social systems, to move out of the laboratory to more ecologically valid situations and to study groups using social VR for longer periods of time.
\end{abstract}

\keywords{virtual reality, 3D user interfaces}



\ifdefined\ARXIV
\else
\maketitle

\fi

\section{Introduction}

Social virtual reality (SVR) applications have been around almost as long as virtual reality systems. For example, the seminal Reality Built for Two system from VPL Research dates from the late 1980s and the machines from Virtuality Systems were available to the general public in arcades from the 1990s. Since these early systems, it has been noted that the immersive nature of systems  creates a collaborative situation that is quite different than video, text and audio-based communication. This appears to be because each user is represented as an avatar and that avatar is dynamically animated by tracking information. Thus body language is quite natural and we can observe spatial interactions between people that can use rules and styles that are inherited from real social situations (e.g. see \cite{schroeder_being_2010}).

There are now a plethora of SVR systems to engage with, many of which are available on consumer VR equipment \cite{mcveigh-schultz_whats_2018, schulz_comprehensive_2020}. There are many different design choices around the avatar representations, capabilities for interaction, tasks supported and social structures that are supported (e.g. see surveys \cite{jonas_taxonomy_2019,Kolesnichenko2019UnderstandingEcology}).

While there is a lot of design variation, prior work of our own group has focused on interaction of small groups. The core activity is that a small group (2-5 persons) are engaged in a task that involves talk around a small number of objects in the environment.  Our concern has been how effective  tasks can be, especially when compared to a group engaging in a similar task in the real world (e.g. from \cite{tromp_small_1998} through to \cite{Moustafa2018ALS}). In this position paper we discuss the trajectories of some of this work. 

\section{Virtual versus Real Interactions}

Something seemingly unique happens when users are immersed in VR: they can be observed using behaviours or showing reactions, that would be appropriate in analogous real-world situations. Thus, even very early SVR systems solicited socially appropriate behaviours from participants, such as offering and receiving of handshakes, social group formation, and other non-verbal behaviours. Such behaviours demonstrate that the users are engaging with others, and many studies have looked at the necessary conditions for such engagement (e.g. see review \cite{oh_systematic_2018}).

One tactic we have used in the past is simply to compare group collaboration in real and virtual worlds. The benchmark being that performance, or ratings of collaboration, should be similar in immersive SVR compared to real world environments (e.g. \cite{steed_leadership_1999, schroeder_collaborating_2001})). While realistic behaviours are still a big concern of ours (see next section), we are developing a theme of work around how users adapt to use of the system over hours and weeks of use \cite{heldal_successes_2005, Moustafa2018ALS}. A key point here has been the participant variability in adaptation to being represented as an avatar and interacting with representations of others, especially those they know well.




\section{Evaluation of communication social cues}

The success in completion of tasks in social VR is a testament to the effectiveness of certain types of non-verbal communication (e.g. \cite{pan_impact_2017}). We can delve into these behaviours to get more detail, but our own experience is that movement is well represented. Obviously more tracking points help (e.g. even foot tracking has an impact \cite{pan_how_2019}). 

Tanenbaum et al. inventory  expressive nonverbal communication in commercial social VR platforms with four categories : movement and proxemic spacing, facial control, gesture and posture and finally virtual environment specific non-verbal communication \cite{tanenbaum_svr_2020}. They note that faces and eye-gaze are a particular remaining problem.  We have tackled eye-gaze in a variety of projects in the past (e.g. \cite{steptoe_eye-tracking_2008, steptoe_lie_2010}). We have found eye-tracking to be highly communicative, but hard to simulate well. Fortunately eye-tracking is a feature we should expect in many more HMDs in the near future. Our own trajectory of work here is to focus on the fine-grained actions that combine multiple verbal and non-verbal acts, to see if they can support certain communicative acts, such as conversation hand-over.

\section{Evaluation of trust}

A particular component of social interaction that we have focused on is trust. We feel that it is useful as a measure because it highlights a slightly higher level of response that the immediacy of gesture or communicative acts, but it is amenable to study in controlled experiments.  Trust is term used in many different ways, but we will refer to behavioural responses, although we note that Salanitri et al. show that the best-known questionnaire to measure usability and trust could be applied to VR \cite{salanitri_trust_2015}. They also showed there is a relationship between user’s satisfaction and trust in the use of VR, and a relationship between usability and trust for different VR technologies. 

Behavioural responses that depend on trust are interesting because they are studied in business and economic contexts. We have adopted a couple of techniques from those fields in our SVR studies: trust through advice seeking (e.g. \cite{pan_comparison_2016}), and trust through a game with stakes that depend on implicit trust of the other(s) \cite{pan2017impact}).  Of course, when embodied by virtual avatars, users might act differently than in real life \cite{junglas_2007}, but we can certain use such methods to study any bias between different communication configurations. This leads to our current objectives in this area: to understand how particular configurations of the VR, such as avatars representation and interface usability bias formation of these higher-level attitudes and relationships between users.







\ifdefined\ARXIV
\bibliographystyle{unsrt}
\else
\bibliographystyle{ACM-Reference-Format}
\fi
\bibliography{sample-base}



\end{document}